\documentclass[aps,pra,twocolumn,preprintnumbers,amsmath,amssymb,superscriptaddress,longbibliography]{revtex4-1}

\input{BellInequalityOptomechanicsPaper.rty}
\nocite{chaturvedi_time-convolutionless_1979,gardiner_adiabatic_1984,wilson-rae_cavity-assisted_2008}

\graphicspath{ {./images/} {./plots/} }

\begin{document}

\title{Violation of Bell's Inequality in Electromechanics}

\author{Sebastian G.\ Hofer}

\affiliation{Vienna Center for Quantum Science and Technology (VCQ), Faculty of Physics, University
  of Vienna, Boltzmanngasse 5, 1090 Vienna, Austria}
\affiliation{Institute for Theoretical Physics,
  Institute for Gravitational Physics (Albert Einstein Institute), Leibniz University Hannover,
  Callinstra\ss{}e 38, 30167 Hannover, Germany}

\author{Konrad W.\ Lehnert} \affiliation{JILA, National Institute of Standards and Technology and
  the University of Colorado, Boulder, CO 80309, USA}
\affiliation{Department of Physics, University of Colorado, Boulder, CO 80309, USA}

\author{Klemens Hammerer} \affiliation{Institute for Theoretical Physics, Institute for
  Gravitational Physics (Albert Einstein Institute), Leibniz University Hannover, Callinstra\ss{}e
  38, 30167 Hannover, Germany}
%
\date{\today}
\begin{abstract}
  Opto- and electromechanical systems offer an effective platform to test quantum theory and its predictions at macroscopic scales. To date, all experiments presuppose the validity of quantum mechanics, but could in principle be described by a hypothetical local statistical theory. Here we suggest a Bell test using the electromechanical Einstein-Podolski-Rosen entangled state recently generated by Palomaki et al.~\cite{Palomaki2013}, which would rule out any local and realistic explanation of the measured data without assuming the validity of quantum mechanics at macroscopic scales. It additionally provides a device-independent way to verify electromechanical entanglement. The parameter regime required for our scheme has been demonstrated or is within reach of current experiments.
\end{abstract}

\maketitle

\emph{Introduction}.---The interaction of light with a mechanically compliant mirror was at the heart of a number of Gedankenexperiments in the early days of quantum theory, and still represents a textbook paradigm illustrating the basic principles and intrinsic limitations of the measurement process in quantum mechanics \cite{Braginsky1995,Aspelmeyer2014}. In recent years experiments in the field of opto- and electromechanics managed to approach some of these textbook examples of controlled quantum dynamics to an amazing degree: Sideband cooling of mechanical oscillators close to the ground state \cite{Teufel2011,chan_laser_2011}, measurement back-action noise \cite{murch_observation_2008,purdy_observation_2013}, ponderomotive squeezing of light \cite{Brooks2012,Safavi-Naeini2013}, coherent quantum state transfer \cite{oconnell_quantum_2010,Palomaki2013a}, feedback control within the thermal decoherence time \cite{wilson_measurement_2014}, and the generation of Einstein-Podolski-Rosen entangled states \cite{Palomaki2013} have all been realized with nano- to micron-sized mechanical oscillators coupled to optical or microwave fields. All of these experiments impressively demonstrate the principles of quantum mechanics working at astonishingly macroscopic scales.

Although all of these effects perfectly confirm the predictions of quantum theory, it is important to note that the data taken in the corresponding experiments could be perfectly well explained in terms of a hypothetical local statistical theory, that is, by a theory assuming local hidden variables that fundamentally determine the ostensibly random measurement results observed in an actual experiment. The possibility of such an alternative, deterministic and local explanation can be ruled out by performing specially designed tests such as the violation of a Bell inequality \cite{Brunner2014}. In their simplest form Bell inequalities present constraints on correlation functions from measurements on the two spatially-separated halves of a bipartite system, which result from the very assumption of local hidden variables. By making measurements on an entangled quantum system a violation can be achieved, thereby excluding any explanation of the observed data based on a local hidden variable theory. Vice versa, the violation of a Bell inequality guarantees the existence of entanglement in a measurement-device-independent manner \cite{bancal_device-independent_2011}.

Here we propose a scheme to violate a Bell inequality using an electromechanical system, which enables us to test local hidden variable models at macroscopic scales. The violation can be achieved with the Einstein-Podolski-Rosen entangled state  recently reported in \cite{Palomaki2013} among a mechanically compliant capacitor and a microwave pulse. In this electromechanical setup high-efficiency non-number resolving detectors for photons and, indirectly, also for phonons can be realized by coupling to a superconducting qubit, as illustrated below. These tools suffice to violate a Bell inequality of the Clauser-Horne-Shimony-Holt (CHSH) type originally introduced in \cite{Banaszek1999}, which was previously violated with two-mode squeezed optical fields \cite{kuzmich_violation_2000}. We show that a significant Bell violation can be attained with parameters that are close to the values from \cite{Palomaki2013} taking into account the most dominant channels of loss and decoherence (\eg{}, cavity losses, photon loss in transmission lines, thermal decoherence of the mechanical oscillator) non-perturbatively.

In view of the recent experiments confirming predictions of quantum theory for nano- and micromechanical systems we suggest, much in line with \cite{asadian_probing_2014}, to strive now for the next level of tests of quantum mechanics challenging classical assumptions of realism and locality at macroscopic scales.

In the following we will first introduce the specific type of Bell inequality relevant to our scheme, then we explain the experimental setup and the protocol for violating this Bell inequality, and finally we present a detailed quantitative model from which we infer prediction for the degree of Bell violation that can be expected under realistic conditions.

\emph{Bell Inequality}.---Let $\bbsigma_{A}(\alpha)$ and $\bbsigma_{B}(\beta)$ be a set of observables for two quantum systems $A$ and $B$, labeled by measurement settings $\alpha$ and $\beta$. Each observable has possible measurement outcomes $\pm1$. Under the premisses of realism and locality the correlations $E_{\alpha\beta}=\mean{\bbsigma_A(\alpha)\otimes\bbsigma_B(\beta)}$ between pairs of measurements for two settings $\alpha_{1(2)}$ and $\beta_{1(2)}$ obey the Bell (CHSH) inequality~\cite{Brunner2014}
\begin{align}\label{eq.BellIneq}
  2\geq |E_{\alpha_1\beta_1}+E_{\alpha_1\beta_2}+E_{\alpha_2\beta_1}-E_{\alpha_2\beta_2}|=:S.
\end{align}
With a suitable choice of observables $\bbsigma_{A}(\alpha)$ and $\bbsigma_{B}(\beta)$ measured on appropriate entangled quantum states this bound may be violated. The maximal violation allowed by quantum mechanics is $S=2\sqrt{2}$ \cite{cirelson_quantum_1980}.

Here we consider measurements on continuous-variable systems performed with a detector that allows to distinguish the vacuum state from all other Fock states. For a mode of an optical field this corresponds, for example, to a standard single-photon counter. If a coherent amplitude $-\alpha$ is added to the mode before it is detected, the measurement effectively  distinguishes between elements of the positive-operator-valued-measure (POVM) $\{P_{\alpha},\mathds{1}-P_{\alpha}\}$, where $P_{\alpha}=\proj{\alpha}$ denotes the projection operator onto the coherent state $\ket{\alpha}$. The technique of measuring this POVM is commonly referred to as weak field homodyning \cite{Wallentowitz1996,Donati2014}. Let the detection of $\ket{\alpha}$ correspond to the measurement result $+1$ and the complementary event to $-1$; the observable is then effectively described by
\begin{equation}
  \label{eq.sigma}
  \bbsigma(\alpha)=\ket{\alpha}\bra{\alpha}-\big[\mathds{1}-\ket{\alpha}\bra{\alpha}\big]=2\ket{\alpha}\bra{\alpha}-\mathds{1}.
\end{equation}
The correlation functions can thus be expressed as $E_{\alpha\beta}=4\mean{P_{\alpha}\otimes P_{\beta}}-2\mean{P_{\alpha}\otimes \mathds{1}+\mathds{1}\otimes P_{\beta}}+1$.
Observables of the form \eqref{eq.sigma} and the corresponding Bell inequality \eqref{eq.BellIneq} have been introduced first in debates regarding the nonlocal properties of spatial superpositions of single photons \cite{Tan1991,Hardy1994,Banaszek1999}, and have been realized experimentally in \cite{Hessmo2004}. Remarkably, the Bell inequality \eqref{eq.BellIneq} can be violated not only with non-Gaussian states (such as single photon states), but even with Gaussian entangled states. For example, a two mode squeezed state
\begin{equation}
  \label{eq.EPRstate}
  \ket\tms_{AB}=\sech r\sum_n\left(-e^{i\varphi}\tanh r\right)^n\ket{n}_A\otimes\ket{n}_B
\end{equation}
yields a maximal violation of $S\approx2.45$ for $6.3$ dB of squeezing ($r\approx0.76$) for optimized values $\alpha_{1(2)}$ and $\beta_{1(2)}$ \cite{Banaszek1999,Brask2012,Lee2009}. An experimental demonstration with squeezed light was reported in \cite{kuzmich_violation_2000}.

An electromechanical two-mode squeezed state has recently been realized in \cite{Palomaki2013}. In the following we will show that this EPR-entangled state shared between a micron-sized mechanical object and a travelling-wave microwave pulse can be used to violate the Bell inequality \eqref{eq.BellIneq}. To perform a measurement of \eqref{eq.sigma} on the electromagnetic mode, we employ a qubit integrated into a microwave cavity, which can directly be used as a single-\emph{photon} detector. In order to achieve single-\emph{phonon} detection on the other hand, the mechanical state has to be first transferred to the microwave field. Assuming the mechanical state is faithfully transferred, another qubit then effectively acts as a single-phonon detector. In this way \emph{photon-phonon} correlations between a microwave pulse and the mechanical oscillator can be inferred from \emph{photon-photon} correlations between two pulses.

\begin{figure}[b]
  \centering
  \includegraphics[width=\columnwidth]{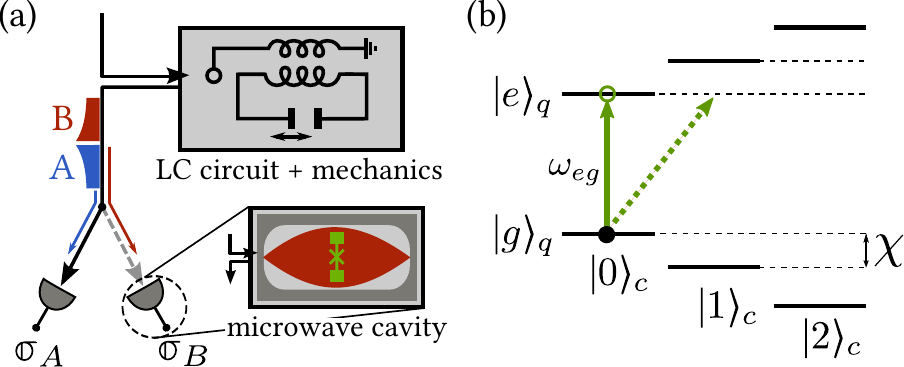}
  \caption{(a) Electromechanical circuit (formed by the LC resonator and the mechanical oscillator) and cascaded microwave cavities with integrated qubits, used as single-photon detectors. (b) Level scheme of superconducting qubit coupled to the intracavity microwave field. The solid green line represents the resonant transition which we seek to drive, while the dashed line indicates an off-resonant transition.}
  \label{fig:setup}
\end{figure}

\emph{The Protocol}.---The proposed protocol can be summarized as follows [see \fref{fig:setup}(a)]. We first generate electromagnetic EPR entanglement between the mechanical mode and a microwave pulse ($A$), by driving the electromechanical system on the blue sideband. The mechanical state is then swapped to a second pulse ($B$) by employing a red-detuned drive \footnote{Note that the center frequency of the scattered light coincides with the cavity resonance frequency.}. Using two microwave cavities containing qubits we can subsequently measure the observables $\bbsigma_A(\alpha)$, $\bbsigma_B(\beta)$ on the two pulses and correlate the measurement results. The protocol thus effectively consists of three steps, which we first discuss in an idealized scenario. Perturbative dynamics will be discussed afterwards. In the following we denote by $c_i$, $c_i^{\dagger}$ the bosonic annihilation and creation operators obeying $[c_i,c_j^{\dagger}]=\delta_{ij}$ (where $i,j$ label subsystems as detailed below).

\pemph{i}~Entanglement Generation: The mechanical oscillator is pre-cooled to its quantum ground state by passive sideband cooling as demonstrated in \cite{Teufel2011}. The electromechanical system is then driven by a blue-sideband pulse for a time $\tau_1$, generating the entangling, two-mode squeezing dynamics $H_{\mathrm{sq}}=g_{\mathrm{sq}}(\clc \cm+\clc^\dagger \cm^\dagger)$ between the mechanical mode ($\m$) and the LC mode ($\slc$) \cite{Palomaki2013}. For an electromechanical coupling $g_{\mathrm{sq}}$ much smaller than the (energy) decay rate $\kappa_{\slc}$ of the LC circuit (weak-coupling regime), entangled photons leave the cavity faster than they are created, such that the mechanical oscillator becomes entangled with a travelling microwave pulse with an exponentially growing temporal profile $\propto\exp(\Gamma_{\mathrm{sq}} t)$, where $\Gamma_{\mathrm{sq}}=4g_{\mathrm{sq}}^2/\kappa_{\slc}$ \cite{hofer_quantum_2011}. The mechanical oscillator and the light pulse will then approximately take on a two-mode squeezed state of the form \eqref{eq.EPRstate}, with $r=\Gamma_{\mathrm{sq}}\tau_1$.

\pemph{ii}~Photodetection: An observable of type \eqref{eq.sigma} can be measured on the microwave pulse as follows. We assume a superconducting qubit is integrated into a cascaded microwave cavity ($\sc$). It is initialized in its ground state $\ket{g}$, and exhibits a switchable dispersive interaction \cite{wallraff_strong_2004,hoffman_coherent_2011,chen_qubit_2014} $H_{q}=\chi(t)\sigma_z \cc^\dagger \cc$ with the cavity, where $\sigma_z$ is the Pauli $z$-matrix. After step \pemph{i}, assuming that the cascaded cavity possesses a bandwidth sufficiently large to accommodate the exponential pulse, the pulse enters the cavity. If at this point the qubit-cavity interaction is switched on, the qubit transition frequency is shifted by $\chi n_{\sc}$ for $n_{\sc}$ cavity photons. Given that the shift per photon $\chi$ is larger than the linewidth of both the qubit and the cavity (this shift can approach values of 1000 times the linewidths of qubit and cavity \cite{paik_observation_2011}), then the qubit can be flipped selectively from its ground to its excited state by applying an external $\pi$-pulse that is resonant within the $n_{\sc}=0$ subspace [see \fref{fig:setup}(b)]. A subsequent measurement of the qubit thus allows us to distinguish the vacuum from all other Fock states. Adding a coherent displacement of $-\alpha$ to the transmission line before the cascaded cavity (or an appropriate amplitude to the cavity directly), results in effectively measuring the observable \eqref{eq.sigma}.

\pemph{iii}~Phonon-Detection: In order to perform a measurement of \eqref{eq.sigma} on the mechanical oscillator, its quantum state is swapped to a second microwave pulse and the measurement of step \pemph{ii} is repeated. The state swap can be achieved by driving the electromechanical system with a red-sideband pulse, which generates a beam-splitter like interaction $H_{\mathrm{bs}}=g_{\mathrm{bs}}(\clc^\dagger \cm+\clc \cm^\dagger)$ \cite{Palomaki2013a}. These dynamics create a microwave pulse whose quantum state ideally is identically to the state of the mechanical oscillator at the end of step \pemph{ii} \cite{hofer_quantum_2011}. However, due to its exponentially decaying temporal envelope $\propto\exp(-\Gamma_{\mathrm{bs}} t)$ (with $\Gamma_{\mathrm{bs}}=4g_{\mathrm{bs}}^2/\kappa_{\slc}$), the pulse will be absorbed by the cavity rather poorly. In order to avoid the associated photon loss we require that both the strength of beam-splitter coupling $g_{\mathrm{bs}}(t)$ and the cascaded cavity's linewidth $\kappa_{\sc}(t)$ can be tuned as a function of time. The coupling strength can be tuned by tailoring the intensity of the drive field incident on the LC circuit; a time-dependent coupling between cavities and transmission lines (\ie{}, a time-dependent linewidth) has recently been demonstrated in \cite{kerckhoff_tunable_2013,pierre_storage_2014,srinivasan_time-reversal_2014,pechal_microwave-controlled_2014,yin_catch_2013,flurin_superconducting_2015}. For an optimized control sequence (see \aref{sec:optimal-pulse-shape} for details) the resulting dynamics approximates an ideal quantum state transfer from the mechanical oscillator to the cavity. Adding an appropriate coherent amplitude $-\beta$ to the microwave pulse thus provides an effective measurement of \eqref{eq.sigma} on the phonon mode via a measurement performed on the microwave pulse.

From the statistics obtained by repeating steps \pemph{i} to \pemph{iii} for fixed amplitudes $\alpha$ and $\beta$ one can compute the correlation $E_{\alpha\beta}$ between the two pulses, which represent \emph{photon-phonon} correlations between the first microwave pulse and the mechanical oscillator. Performing the procedure for appropriate amplitudes (measurement settings) $\alpha_{1(2)}$ and $\beta_{1(2)}$ ultimately allows to  violate the Bell inequality \eqref{eq.BellIneq}.

\emph{The Model}.---To show that a violation of a Bell inequality can be achieved in state-of-the-art electromechanical experiments \cite{Palomaki2013a,Palomaki2013} we provide a detailed model of all steps, including the dominant decoherence channels. In particular we include non-perturbatively mechanical decoherence, photon losses, and counter-rotating terms of the radiation-pressure interaction. In order to model the measurement of \eqref{eq.sigma} using one of the microwave cavities containing a qubit, we treat the cavity as a cascaded system \cite{gardiner_driving_1993,carmichael_quantum_1993}, to which the LC circuit couples unidirectionally. This allows us to correctly describe the transfer of the pulse into the cavity without treating it explicitly. The state of the three modes (mechanics, LC circuit, one of the microwave cavities) is described by the density matrix $\mu$, whose evolution during steps \pemph{i} and \pemph{iii} follows the master equation
\begin{multline}
  \label{eq.meq}
  \dot\mu=-i[\omega_m\cm^\dagger \cm-\Delta \clc^\dagger \clc+(g\clc+g^*\clc^\dagger)(\cm+\cm^\dagger),\mu]\\
  +\mathcal{L}_m\mu
  +\kappa_{\slc}\mathcal{D}[\clc]\mu+\kappa_{\sc}\mathcal{D}[\cc]\mu\\
  -\sqrt{\lambda_t\kappa_{\slc}\kappa_{\sc}/4}\left\{[\cc^\dagger,\clc\mu]+[\mu \clc^\dagger,\cc]\right\}.
\end{multline}
In the Hamiltonian dynamics (first line) $\omega_m$ denotes the mechanical frequency, $\Delta=\omega_{\mathrm{drive}}-\omega_{\slc}$ the detuning between the frequencies of the LC mode $\omega_{\slc}$ and the drive field $\omega_{\mathrm{drive}}$, and $g(t)=\frac{g_0}{\kappa_{\slc}/2-i\Delta}\sqrt{P(t)\kappa_{\slc}/2\hbar\omega_{\slc}}$ is the linearized optomechanical coupling. The coupling per single photon is denoted by $g_0$ and $P(t)$ is the  power of the drive field which may vary slowly in time as long as $\dot{P}/P\ll\max(\kappa_{\slc},|\Delta|)$ \cite{hofer_quantum_2011}. The second line describes decoherence processes by means of Lindblad operators $\mathcal{D}[a]\mu=a\mu a^\dagger-\frac{1}{2}a^\dagger a\mu-\mu a^\dagger a$ and $\mathcal{L}_m\mu=\gamma_m(\bar{n}+1)\mathcal{D}[\cm]\mu+\gamma_m\bar{n}\mathcal{D}[\cm^{\dagger}]\mu$. The full-width-at-half-maximum damping rate of the mechanical oscillator is $\gamma_m$ and $\bar{n}=[\exp(\hbar\omega_m/k_BT)-1]^{-1}$ is its mean occupation number in thermal equilibrium at temperature $T$. The third line models the cascaded coupling of the electromagnetic system into the microwave cavity \cite{gardiner_driving_1993,carmichael_quantum_1993}. The efficiency of the transmission channel is $\lambda_t$.

The master equation \eqref{eq.meq} describes Gaussian dynamics, and can in principle be integrated exactly. In order to speed up integration and numerical optimization, we adiabatically eliminate the LC circuit (valid in the weak-coupling regime $g\ll \kappa_{\slc}$) and integrate the dynamics in a frame rotating at the mechanical frequency $\omega_m$, which is by far the fastest time scale in the problem (see \aref{sec:schrodinger-picture}). In step \pemph{i} the detuning is chosen on the first blue sideband, $\Delta=\omega_m$, yielding the effective master equation
\begin{equation}
  \label{eq:17}
  \begin{split}
    \dot{\rho} &= \mathcal{L}_{m} \rho + \epsilon \Gamma_{\mathrm{sq}}\mathcal{D}[\cm]\rho+ \mathcal{D}[\sqrt{\kappa_{\sc}}\cc-i\sqrt{\lambda_t\Gamma_{\mathrm{sq}}}\cm^{\dagger}]\rho\\
    &+(1-\lambda_t)\Gamma_{\mathrm{sq}}\mathcal{D}[\cm^{\dagger}]\rho + \frac{i}{2} \sqrt{\lambda_t\Gamma_{\mathrm{sq}}\kappa_{\sc}}\,[\cm\cc+\cm^{\dagger}\cc^{\dagger},\rho]
  \end{split}
\end{equation}
for $\rho=\ptr{\slc}{\mu}$, with $\epsilon=1/[1+(4\omega_m/\kappa_{\slc})^2]$. As it turns out, it is advantageous to slightly mismatch the cavity's bandwidth with respect to the exponential envelop of the light pulse; this is due to the finite duration of the pulse, which causes a spectral broadening. We therefore set $\kappa_{\sc}=\upsilon\Gamma_{\mathrm{sq}}$, and optimize later with respect to $\upsilon$. In step \pemph{iii} we use a red-detuned pulse with $\Delta=-\omega_m$, leading to the equation
\begin{equation}
  \label{eq:18}
  \begin{split}
    \dot{\rho} &= \mathcal{L}_{m} \rho + \epsilon \Gamma_{\mathrm{bs}}\mathcal{D}[\cm^{\dagger}]\rho+ \mathcal{D}[\sqrt{\kappa_{\sc}}\cc-i\sqrt{\lambda_t\Gamma_{\mathrm{bs}}}\cm]\rho\\
    &+(1-\lambda_t)\Gamma_{\mathrm{bs}}\mathcal{D}[\cm]\rho + \frac{i}{2} \sqrt{\lambda_t\Gamma_{\mathrm{bs}}\kappa_{\sc}}\,[\cm^{\dagger}\cc+\cm\cc^{\dagger},\rho].
  \end{split}
\end{equation}
Both the linewidth $\kappa_{\sc}(t)$ of the cavity and the amplitude $\sqrt{P(t)}$ of the pulse [and therefore the effective electromechanical coupling strength $\Gamma_{\mathrm{bs}}(t)$] needs to be shaped as detailed in \aref{sec:optimal-pulse-shape} to maximize the read-out efficiency.

In order to evaluate the quantity $S$ in \eqref{eq.BellIneq} for the bipartite system consisting of the two light pulses,
we in turn integrate equations \eqref{eq:17} and \eqref{eq:18} for durations $\tau_1$ and $\tau_2$ respectively, assuming that initially  the mechanical mode is in a thermal state with a mean occupation number $n_0$ and the respective microwave cavity is in the vacuum state right before the arrival of the pulse. As the system is Gaussian, its state is fully determined by the first and second moments of the vector $\vc{X}=(x_{\m},y_{\m},x_{\sc},y_{\sc})$, where $x_k$ and $y_k$ are quadrature operators obeying $[x_k,y_l]=i\delta_{kl}$. To evaluate the quantity $S$ it suffices to calculate the symmetrized covariance matrix $\Sigma_{kl}=\frac{1}{2}\mean{X_kX_l+X_lX_k}-\mean{X_k}\mean{X_l}$ at the end of the pulse sequence. The master equations \eqref{eq:17} and \eqref{eq:18} lead to a differential \emph{Lyapunov} equation of the form  \(\dot{\mat{\Sigma}}=\mat{F}\mat{\Sigma}+\mat{\Sigma}\mat{F}^{\trans}+\mat{N}\) \cite{edwards_optimal_2005}.
The explicit form of the matrices $\mat{F}$ and $\mat{N}$ is given in \aref{sec:gaussian-dynamics}. The Lyapunov equation is linear and can be integrated analytically [even for time-dependent parameters $\Gamma_{\mathrm{bs}}(t)$, $\kappa_{\sc}(t)$]. The covariance matrix $\mat{\Sigma}$ determines the characteristic function from which the Bell inequality violation \eqref{eq.BellIneq} can be calculated along the lines of \cite{Banaszek1999,Lee2009}.

\begin{figure}[tb]
  \includegraphics[width=\columnwidth]{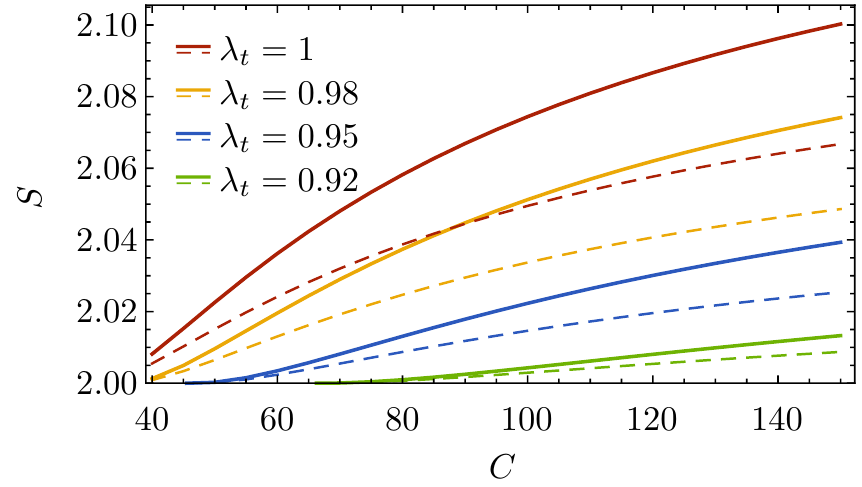}
  \caption{Bell correlations against cooperativity optimised with respect to $\tau_1$, $\tau_2$ and $\upsilon$ for different transmissivities $\lambda_t=1,0.98,0.95,0.92$ (red, yellow, green, blue), and initial mechanical occupation numbers $n_0=0.1,0.25$ (solid, dashed lines). Other parameters are $\bar{n}=40$ and $\kappa_{\slc}/\om\approx 1/8$.}
  \label{fig:bellcorrelation}
\end{figure}

\emph{Results}.---We optimize the resulting value of $S$ with respect to the measurement settings $\alpha_{1(2)}$ and $\beta_{1(2)}$, the pulse duration $\tau_1$ and $\tau_2$, and linewidth of the microwave cavity [in step \pemph{i} only], parameterized by $\upsilon$ as discussed above. The optimization is performed for a fixed transmission loss $1-\lambda_t$, bath temperature $T$ and for a given maximal coupling $g_{\mathrm{max}}=\sup_t g(t)$. To facilitate the comparison between different experimental platforms, it is instructive to parameterize this coupling strength by means of the cooperativity $C=4g_{\mathrm{max}}^2/\kappa_{\slc}\ga(\bar{n}+1)$. The results for the maximal Bell correlations are plotted in \fref{fig:bellcorrelation} versus $C$, for different values of the transmissivity $\lambda_t$ and the initial mechanical occupation number $n_0$.

\begin{figure}[tb]
  \includegraphics[width=\columnwidth]{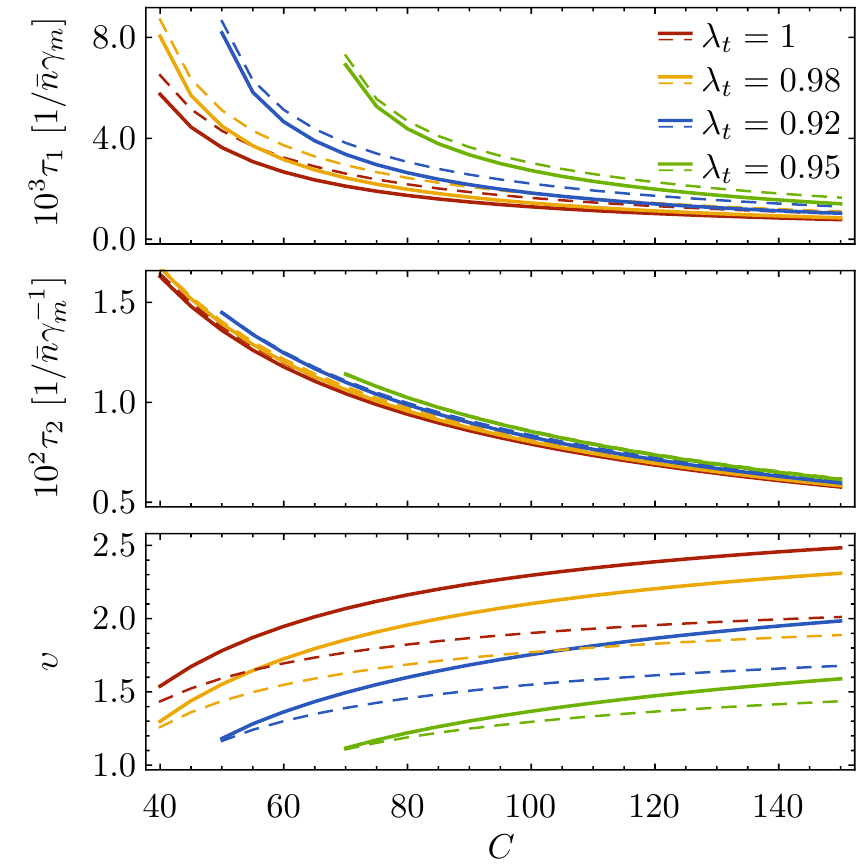}
  \caption{Optimal values of $\tau_1$ (top), $\tau_2$ (middle), and $\upsilon$ (bottom) for the corresponding values of $S$ in \fref{fig:bellcorrelation}.}
  \label{fig:pulselength}
\end{figure}

We conclude that a significant violation of the Bell inequality \eqref{eq.BellIneq} can be achieved with cooperativities and initial mechanical occupation numbers that are feasible in electromechanical systems. Cooperativity values of up to $C\approx 300$ and occupation numbers of $n_0=0.34$ and $0.25$ have, for example, been demonstrated in \cite{Teufel2011} and \cite{lecocq_resolving_2015}, respectively. The greatest challenge will be to bring the overall transfer efficiency above 90\%. This is a lively research activity in the superconducting qubit community, however.

The optimal values for $\tau_1,\,\tau_2$ and $\upsilon$ maximizing the Bell violation are shown in \fref{fig:pulselength}, from which we infer that the optimal values of both $\tau_1$ and $\tau_2$ lie well below the effective coherence time $1/\bar{n}\gamma_{\m}$ of the mechanical system. For increasingly long pulses (for low values of $C$), the optimal value for the decay rate of the microwave cavity is $\kappa_{\sc}\approx \Gamma_{\mathrm{sq}}$, as expected.

\emph{Conclusion}.---In this Letter we present an effective scheme to demonstrate the violation of Bell's inequality using EPR entanglement shared between a mechanical oscillator and a microwave field. We analyse in detail the experimental implementation, including the primary decoherence channels, such as photon losses and thermal mechanical noise. We show that a significant violation of Bell's inequality is achievable with electromechanical systems. We want to emphasize that using an experimentally considerably less complex setup employing only a single detection setup can still be used to demonstrate electromechanical entanglement in a device-independent manner.

Note that an equivalent scheme can be considered in the optical domain using conventional photodetectors instead of the qubit as a photon counter. During preparation of this manuscript we became aware of related work along this line by Vivoli et al.~\cite{vivoli_optomechanical_2015}.
\appendix
\section{Photodetection}
\label{sec:photodetection}

A measurement corresponding to the POVM $\{\ket{\alpha}\bra{\alpha},\mathds{1}-\ket{\alpha}\bra{\alpha}\}$ on the cavity mode $\sc$ can be realized in the following way: Due to the dispersive interaction, the intracavity field shifts the qubit's resonance frequency by $n\chi$, where $n$ is the number of intracavity photons. We start with the qubit in the ground state and apply a $\pi$-pulse at a frequency $\oqb$. If the energy levels of the qubit are well resolved, \ie{}, the frequency shift between adjacent levels $\chi$ is larger than the linewidth of the qubit's excited state, we can describe the unitary evolution generated due to the $\pi$-pulse by
\begin{multline}
  \label{eq:10}
  U_{\pi}= \ket{0,e}\bra{0,g}_{\sc q}+\ket{0,g}\bra{0,e}_{\sc q}+\sum_{l=1}^{\infty}\proj{l}_{\sc} \otimes \mathds{1}_q
\end{multline}
Measuring the state of the qubit after the $\pi$-pulse is then described by the measurement operators $M_r={\bra{r}}U_{\pi}\ket{g}_{q}$ for possible outcomes $r\in \left\{ g,e \right\}$. We thus see that measuring the qubit realizes a projective measurement on the cavity field where $M_e=\proj{0}_{\sc}$, $M_g=\mathds{1}_{\sc}-M_e$ correspond to finding the qubit in the excited and ground state respectively.
The probability to find the outcome $r$ is $p_r=\tr{M_r^{\dagger}M_r\rho}$. If before applying the $\pi$-pulse, we displace the cavity field by $-\alpha$, \ie{}, $\rho\rightarrow D(-\alpha)\rho D^{\dagger}(-\alpha)$ and set $r=e$, we find
\begin{equation}
  \label{eq:21}
  \begin{aligned}
    p_e&=\tr{\proj{0}_{\sc}D^{\dagger}(\alpha)\rho D(\alpha)}\\
    &=\tr{\proj{\alpha}_{\sc}\rho}=\mean{P_{\alpha}},
  \end{aligned}
\end{equation}
which is what we need to evaluate the parameter $S$.

\section{Pulse Shape Optimisation}
\label{sec:optimal-pulse-shape}

\begin{figure}[htbp]
  \includegraphics{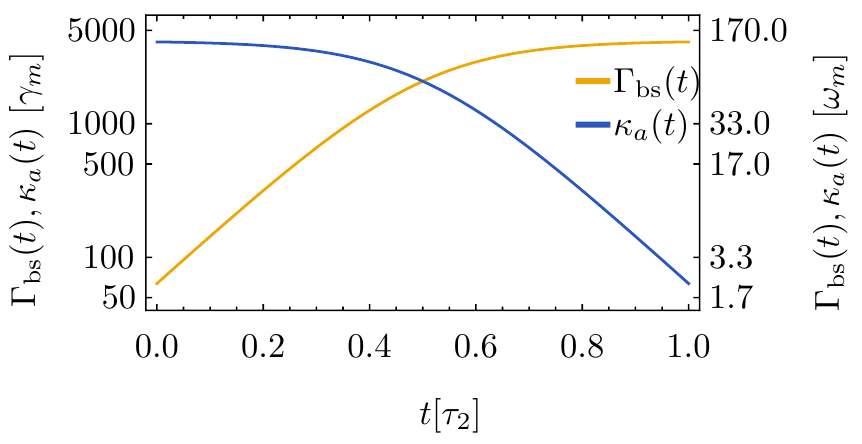}
  \caption[]{Optimal time-dependent value of $\Gamma_{\mathrm{bs}}(t)=4g_{\mathrm{bs}}(t)^2/\kappa_{\slc}$ (yellow line) and $\kappa_{\sc}(t)$ (blue line) for the second pulse in terms of the mechanical FWHM linewidth $\gamma_m$ (right axis) and the mechanical resonance frequency (left axis), where we assumed a Q-factor of $Q=3\cdot 10^6$.}\label{fig:oshape}
\end{figure}

Here we discuss the optimal-control problem of how to transfer the mechanical quantum state into a cascaded cavity. We use a red-detuned ($\Delta=-\om$) light pulse of length $\tau_2$, whose power $P(t)$ can be varied in time [leading to a adiabatic coupling strength $\Gamma_{\mathrm{bs}}(t)$]. Additionally we assume we can tune the bandwidth $\kappa_{\sc}(t)$ of the cascaded cavity. To identify the relevant temporal shapes we rewrite the system's evolution in terms of the adiabatic Langevin equations corresponding to master equation \eqref{eq:18}. Here we are only interested in the classical dynamics and we thus introduce the mean values $\beta(t)=\mean{\cm(t)}$ and $\xi(t)=\mean{\cc(t)}$. Their equations of motion are
\begin{subequations}
  \begin{align}
    \label{eq-ops:1}
    \dot{\beta}(t)&=-\frac{\Gamma_{\mathrm{bs}}(t)}{2}\beta(t),\\
    \label{eq-ops:2}
    \dot{\xi}(t)&=-\frac{\kappa_{\sc}}{2}(t)\xi(t)-\ii \sqrt{\Gamma_{\mathrm{bs}}(t)\kappa_{\sc}(t)}\beta(t).
  \end{align}
\end{subequations}
Clearly \eqref{eq-ops:1} has the solution
\begin{equation}
  \label{eq-ops:4}
  \beta(t)=\exp \left( -\frac{1}{2}\int_0^t\dd{\tau}\Gamma_{\mathrm{bs}}(\tau) \right)\beta(0)
\end{equation}
and we can thus write, after formally integrating \eqref{eq-ops:2},
\begin{multline}
  \label{eq-ops:3}
  \xi(\tau_2)=\ee^{-\frac{1}{2}\int_0^{\tau_2}\dd{t}\kappa_{\sc}(t)}\xi(0)
  -\ii  \Big[\int_0^{\tau_2}\dd{t}\sqrt{\Gamma_{\mathrm{bs}}(t)\kappa_{\sc}(t)}\times\\
  \ee^{-\frac{1}{2}\int_t^{\tau_2}\dd{s}\kappa_{\sc}(s)}\ee^{-\frac{1}{2}\int_0^t\dd{s}\Gamma_{\mathrm{bs}}(s)}\Big] \beta(0).
\end{multline}
Our goal now is to maximize the term in brackets in the second line (we call it $I$) which quantifies the fidelity of the state swap and fulfills $0\leq I\leq 1$. We define two functions
\begin{subequations}
  \label{eq-ops:7}
  \begin{align}
    v(t)&=\sqrt{\Gamma_{\mathrm{bs}}(t)}\ee^{-\frac{1}{2}\int_0^t\dd{s}\Gamma_{\mathrm{bs}}(s)},\\
    w(t)&=\sqrt{\kappa_{\sc}(t)}\ee^{-\frac{1}{2}\int_t^{\tau_2}\dd{s}\kappa_{\sc}(s)},
  \end{align}
\end{subequations}
which we assume to be square integrable. We can thus write the overlap $I$ as a scalar product on the underlying vector space, which obeys the Cauchy--Schwartz inequality
\begin{equation}
  \label{eq-ops:8}
  I=\mean{v,w}^2\leq \mean{v,v}\mean{w,w}.
\end{equation}
The right-hand side is easily evaluated and we find
\begin{align*}
  \mean{v,v}&=1-\ee^{-K_v},\\
  \mean{w,w}&=1-\ee^{-K_w}
\end{align*}
with $K_v=\int_0^{\tau_2}\dd{s}\Gamma_{\mathrm{bs}}(s)$ and $K_w=\int_0^{\tau_2}\dd{s}\kappa_{\sc}(s)$. The inequality \eqref{eq-ops:8} is saturated for the choice $v\equiv w$, or equivalently (as $v,w\geq 0$) for $v^2\equiv w^2$. A possible choice for $\Gamma_{\mathrm{bs}}$ and $\kappa_{\sc}$ is thus
\begin{subequations}
  \label{eq-ops:13}
  \begin{align}
    \Gamma_{\mathrm{bs}}(t)&=N \ee^{-\int_t^{\tau_2}\dd{s}\kappa_{\sc}(s)},\\
    \kappa_{\sc}(t)&=N \ee^{-\int_0^t\dd{s}\Gamma_{\mathrm{bs}}(s)},
  \end{align}
\end{subequations}
where $N$ fixes their norm. The set of differential equations corresponding to \eqref{eq-ops:13} is
\begin{subequations}
  \label{eq-ops:14}
  \begin{align}
    \dot{\Gamma_{\mathrm{bs}}}(t)&=\kappa_{\sc}(t)\Gamma_{\mathrm{bs}}(t),\\
    \dot{\kappa}_{a}(t)&=-\kappa_{\sc}(t)\Gamma_{\mathrm{bs}}(t),
  \end{align}
\end{subequations}
with the boundary conditions
\begin{equation}
  \label{eq:297}
  \Gamma_{\mathrm{bs}}(\tau_2)=\kappa_{\sc}(0)=N.
\end{equation}
Non-singular solutions of these equations are given by
\begin{subequations}
  \label{eq-ops:16}
  \begin{align}
    \Gamma_{\mathrm{bs}}(t)&=\frac{M}{1+\ee^{M(2t-\tau_2)}},\\
    \kappa_{\sc}(t)&=\frac{M}{1+\ee^{-M(2t-\tau_2)}},
  \end{align}
\end{subequations}
where the parameter $M$ is determined by the condition $\Gamma_{\mathrm{bs}}(\tau_2)=\kappa_{\sc}(0)=M(1+\ee^{M\tau_2})^{-1}$. Functions \eqref{eq-ops:16} are shown in \fref{fig:oshape}.

\section{Adiabatic Elimination of the LC Mode}
\label{sec:schrodinger-picture}

\subsection{Time-Convolutionless Projection Operator Method}
Adiabatic elimination of the cavity mode on the basis of the master equation (MEQ) \eqref{eq.meq} can be achieved through the time-convolutionless projection operator method \cite{chaturvedi_time-convolutionless_1979}. Given a set of linear operators $\cL_i$ ($i\in \{0,1\}$) and a corresponding equation of the form
\begin{equation}
  \label{eq-ae:77}
  \frac{\dd{}}{\dt}\mu(t)=\left[\cL_0 + \varepsilon \cL_1(t) +\varepsilon^2 \cL_2 \right]\mu(t),
\end{equation}
the goal is to find the dynamics of $\mu$ on a subspace defined by the projection $\mathcal{P}=\mathds{1}-\mathcal{Q}$. We first transform into an interaction picture with $\cL_0$ and define
\begin{subequations}
  \label{eq-ae:206}
  \begin{align}
    \tilde{\mu}(t) &= \exp{(-\cL_0 t)}\mu(t),\\
    \tilde{\cL}_i(t) &= \exp{(-\cL_0t)}\cL_i(t)\exp{(\cL_0t)}.
  \end{align}
\end{subequations}
The equation of motion for $\mathcal{P}\mu$ can then be written in the form
\begin{equation}
  \label{eq-ae:78}
  \frac{\dd{}}{\dt}\mathcal{P}\tilde{\mu}(t)=\mathcal{K}(t)\mathcal{P}\tilde{\mu}(t),
\end{equation}
where we assumed that $\mathcal{Q}\mu(0)=0$. We are only interested in a expansion of $\mathcal{K}$ in powers of a small parameter $\varepsilon$, which we write as $\mathcal{K}=\sum_n\varepsilon^n\mathcal{K}_n$. We can show that up to second order the expansion coefficients are given by
\begin{subequations}
  \label{eq-ae:225}
  \begin{align}
    \mathcal{K}_1(t)&= \mathcal{P}\tilde{\cL}_1(t)\mathcal{P},\\
    \label{eq-ae:268}
    \mathcal{K}_2(t)&= \mathcal{P}\tilde{\cL}_2(t)\mathcal{P} + \mathcal{P}\tilde{\cL}_1(t)\int_{0}^t \dd{\tau} \mathcal{Q}\tilde{\cL}_1(\tau)\mathcal{P},
  \end{align}
\end{subequations}
and $\mathcal{K}_0=0$. In order to apply this method for adiabatic elimination we need to identify a suitable subspace $\mathcal{P}$ which describes the relevant dynamics. This subspace must be chosen such that \cite{gardiner_adiabatic_1984}
\begin{subequations}
  \label{eq-ae:209}
  \begin{gather}
    \mathcal{P}\cL_0=\cL_0\mathcal{P}=0,\\
    \mathcal{P}\cL_1\mathcal{P}=0,\\
    \mathcal{P}=\lim_{t\rightarrow 0}\ee^{\cL_0t}.
  \end{gather}
\end{subequations}

\subsection{Elimination of the LC Mode}
\label{sec:elim-optic-mode}
To eliminate the {LC} mode we choose $\mathcal{P}\mu=\ptr{\slc}{\mu}\otimes \rho_{\vac}$ which projects the state $\mu$ onto the cavity's ground state. This is the subspace we are interested in, as in the limit $g\ll \kappa_{\slc}$ (in this section we take $g$ to stand for either $g_{\mathrm{sq}}$ or $g_{\mathrm{bs}}$) all photons scattered from the mechanical oscillator into the resonator (typically on a timescale $1/g$) immediately decay from it (on a much shorter timescale $1/\kappa_{\slc}$). For the microwave cavity we have $\kappa_{\sc}\approx 4g^2/\kappa_{\slc}$ and thus $\kappa_{\sc}/\kappa_{\slc}\approx(2g/\kappa_{\sc})^2$. It is convenient to introduce the parameter $\bar{\kappa}=\sqrt{\kappa_{\slc}\kappa_{\sc}}$. Consequently we assume a separation of time scales of the form
\begin{equation*}
  \kappa_{\sc} \ll g,\bar{\kappa}\ll\kappa_{\slc},\om,\Dc,\Dlc
\end{equation*}
which reflects the structure of the Liouvillian if we identify $\varepsilon$ with $g/\kappa_{\sc}$. We neglect the mechanical decoherence for now (\ie{}, we set $\gamma_m=0$) and add it again in the end. (This can be shown to be exact.)

By going into an interaction picture with the free Hamiltonian $\om \cm^{\dagger}\cm-\Dc\cc^{\dagger}\cc$ we can write the {MEQ} \eqref{eq.meq} in the required form $\dot{\mu}=[\cL_0+\cL_1(t)+\cL_2]\mu$ with the definitions
\begin{subequations}
  \begin{gather}
    \label{eq-ae:210}
    \cL_0\mu=-\ii \Delta_{\slc}[\clc^{\dagger}\clc,\mu]+\kappa_{\slc} \lb[\clc]\mu,\\
    \cL_1(t)=\ee^{\ii\om t}\cL_{\m}^+\mu+\ee^{-\ii\Dc t}\cL_{\sc}^-\mu+\Hc{}\\
    \cL_2=\kappa_{\sc}\lb[\cc]\mu,
    \shortintertext{and}
    \cL_{\m}^+\mu=-\ii g[\cm^{\dagger}(\clc+\clc^{\dagger}),\mu],\\
    \cL_{\m}^-\mu=(\cL_{\m}^+\mu)^{\dagger}=-\ii g[\cm(\clc+\clc^{\dagger}),\mu],\\
    \cL_{\sc}^+ \mu = -\sqrt{\lambda_t}\bar{\kappa} [\cc^{\dagger},\clc \mu],\\
    \cL_{\sc}^- \mu = (\cL_{\sc}^-\mu)^{\dagger} = \sqrt{\lambda_t}\bar{\kappa} [\cc,\mu \clc^{\dagger}],
  \end{gather}
\end{subequations}
We can then show equations \eqref{eq-ae:209} and additionally find the useful relations
\begin{subequations}
  \label{eq-ae:212}
  \begin{gather}
    \mathcal{Q}\cL_0 \mathcal{Q} = 0,\\
    \mathcal{P}\ee^{\cL_0 t}=\ee^{\cL_0 t}\mathcal{P}=\mathcal{P},
  \end{gather}
\end{subequations}
which we can use to evaluate equations \eqref{eq-ae:225}. We first introduce \(\rho(t)=\ptr{\slc}{\mu(t)}\) and thus have \(\mathcal{P}\mu(t)=\rho(t)\otimes \rho_{\vac}\). We then immediately find $\mathcal{K}_1(t)=\mathcal{P}\tilde{\cL}_1(t)\mathcal{P}=\mathcal{P}\cL_1(t)\mathcal{P}=0$ and $\mathcal{P}\tilde{\cL}_2\mathcal{P}=\cL_2\mathcal{P}$. The second term in \eqref{eq-ae:268} is more involved. Taking into account that \(\cL_{\sc}^{\pm}\mathcal{P}=0\) we find the expanded expression
\begin{equation*}
  \begin{split}
    \MoveEqLeft\mathcal{P}\tilde{\cL}_1(t)\int_0^t\dd{\tau}Q\tilde{\cL}_1(\tau)\mathcal{P}\rho(t) =\\
    & \int_0^t\dd{\tau} \ee^{\ii \om \tau} \mathcal{P}\cL_{\m}^+ \ee^{\cL_0 \tau}\cL_{\m}^- \mathcal{P}\rho(t)\\
    & + \int_0^t\dd{\tau} \ee^{\ii \om \tau} \mathcal{P}\cL_{\m}^- \ee^{\cL_0 \tau}\cL_{\m}^- \mathcal{P}\rho(t)\ee^{-2\ii \om t}\\
    & + \int_0^t\dd{\tau} \ee^{\ii \om \tau} \mathcal{P}\cL_{\sc}^+ \ee^{\cL_0 \tau}\cL_{\m}^- \mathcal{P}\rho(t)\ee^{-\ii (\Dc+\om) t}\\
    & + \int_0^t\dd{\tau} \ee^{-\ii \om \tau} \mathcal{P}\cL_{\sc}^+ \ee^{\cL_0 \tau}\cL_{\m}^+ \mathcal{P}\rho(t)\ee^{-\ii (\Dc-\om) t}+\Hc{}
  \end{split}
\end{equation*}
The first term results from the optomechanical interaction alone and will give rise to the familiar heating and cooling terms derived in \cite{wilson-rae_cavity-assisted_2008}. The second term is fast oscillating and will later be dropped in a rotating-wave-approximation. The last two terms correspond to scattering of photons into the cascaded cavity mode. Note that depending on the detuning of the second cavity only one of the mechanical sidebands is resonantly scattered into the second cavity mode. We evaluate the important terms separately, treating the integrand only for the moment. The first term gives
\begin{multline*}
  \ptr{\slc}{\mathcal{P}\cL_{\m}^+\ee^{\cL_0\tau}\cL_{\m}^-\mathcal{P}\rho(t)} =\\
  - g^2 \bigl( \ptr{\slc}{\clc(\tau)\clc^{\dagger}\rho_0}[\cm^{\dagger},\cm\mu] - \ptr{\slc}{\clc\clc^{\dagger}(\tau) \rho_0} [\cm^{\dagger},\mu \cm] \bigr).
\end{multline*}
The third and fourth term give respectively
\begin{align*}
  \ptr{\slc}{\mathcal{P}\cL_{\sc}^+\ee^{\cL_0\tau}\cL_{\m}^-\mathcal{P}\rho(t)} &=
    \ii g \sqrt{\lambda_t}\bar{\kappa} \ptr{\slc}{\clc(\tau)\clc^{\dagger}\rho_0}[\cc^{\dagger},\cm\mu],\\
  \ptr{\slc}{\mathcal{P}\cL_{\sc}^+\ee^{\cL_0\tau}\cL_{\m}^+\mathcal{P}\rho(t)} &= \ii g \sqrt{\lambda_t}\bar{\kappa} \ptr{\slc}{\clc(\tau)\clc^{\dagger}\rho_0}[\cc^{\dagger},\cm^{\dagger}\mu].
\end{align*}
As we will drop the second term in a rotating-wave approximation (RWA) we do not state the explicit expression here. Next we need to find the correlation functions \(\ptr{\slc}{\clc(t)\clc^{\dagger}(0)\rho_{\vac}}\) and \(\ptr{\slc}{\clc(0)\clc^{\dagger}(t)\rho_{\vac}}=\ptr{\slc}{\clc(t)\clc^{\dagger}(0)\rho_{\vac}}^{*}\). These can be calculated using the quantum regression theorem. We find
\begin{subequations}
  \begin{align}
    \ptr{\slc}{\clc(t)\clc^{\dagger}(0)\rho_{\vac}} &= \exp{[(\ii \Dlc - \kappa_{\slc}/2)t]},\\
    \ptr{\slc}{\clc(0)\clc^{\dagger}(t)\rho_{\vac}} &= \exp{[-(\ii \Dlc + \kappa_{\slc}/2)t]}.
  \end{align}
\end{subequations}
If we neglect terms of the form $\exp(-\kappa_{\slc}t)$ we thus find
\begin{subequations}
  \begin{align}
    \int_0^t\dd{\tau}\ee^{\pm\ii\om \tau}\ptr{\slc}{\clc(\tau)\clc^{\dagger}(0)\rho_{\vac}} &\approx \eta_{\mp},
  \end{align}
\end{subequations}
where $\eta_{\pm}=[\tfrac{\kappa_{\slc}}{2}-\ii(-\Delta_{\slc}\pm \om)]^{-1}$. Taking this all together we neglect fast rotating terms in a {RWA}, and find
\begin{multline}
  \label{eq-ae:200}
  \dot{\rho}=-\ii[\delta\om \cm^{\dagger}\cm,\rho] +\Gamma_- \lb[\cm]\rho+\Gamma_+ \lb[\cm^{\dagger}]\rho + \kappa_{\sc} \lb[\cc]\rho \hfill\hfill\\
  \hfill-g\sqrt{\lambda_t\kappa_{\sc}/\kappa_{\slc}}\bigr\{ [\cc^{\dagger},s_+\rho]\ee^{-\ii (\Dc-\om)t}+\Hc{} \bigl\}\hfill\\
  -g\sqrt{\lambda_t\kappa_{\sc}/\kappa_{\slc}}\bigr\{ [\cc^{\dagger},s_-\rho]\ee^{-\ii (\Dc+\om)t}+\Hc{} \bigl\},
\end{multline}
where we introduced jump operators $s_+=-\ii \kappa_{\slc}\eta_+\cm^{\dagger}$ and $s_-=-\ii\kappa_{\slc}\eta_-\cm$. The other parameters are given by $\delta\omega_m=g^2\mathrm{Im}(\eta_-+\eta_+)$, $\Gamma_{\pm}=2g^2\mathrm{Re}\eta_{\pm}$. To get a time-independent {MEQ} we first go into an interaction picture with $\delta\om \cm^{\dagger}\cm$, which eliminates the first term in \eqref{eq-ae:200} and leads to the replacement $\ee^{\pm\ii \om t}\rightarrow \ee^{\pm \ii \om^{\smash{\mathrm{eff}}}t}$ (with $\om^{\mathrm{eff}}=\om+\delta\om$). We then introduce an averaged density operator, formally defined by
\begin{equation}
  \label{eq-ae:271}
  \bar{\rho}(t)\dt=\int_t^{t+\delta t}\dd{\tau}\rho(\tau).
\end{equation}
$\delta t$ has to be chosen to be long on the mechanical oscillation frequency, but short on all other timescales. This leads to the inequality $\om^{\mathrm{eff}}\gg 1/\delta t \gg g^2/\kappa_{\slc},g\sqrt{\kappa_{\sc}/\kappa_{\slc}}$. If we now consider the cases $\Dc=\pm \om^{\mathrm{eff}}$ and neglect fast rotating terms [which give a correction of order \(\order(1/\om^{\mathrm{eff}} \delta t)\)] we eventually obtain for $\bar{\rho}$ the coarse-grained {MEQ}
\begin{multline}
  \label{eq-ae:270}
  \dot{\bar{\rho}}=  \Gamma_- \lb[\cm]\bar{\rho}+\Gamma_+ \lb[\cm^{\dagger}]\bar{\rho} + \kappa_{\sc} \lb[\cc]\bar{\rho}\\
  - g\sqrt{\lambda_t\kappa_{\sc}/\kappa_{\slc}}([\cc^{\dagger},s_{\pm}\bar{\rho}]+[\bar{\rho} s_{\pm}^{\dagger},\cc]).
\end{multline}
We can now use the identity
\begin{equation*}
  \mathcal{D}[a+b]\rho=\mathcal{D}[a]\rho+\mathcal{D}[b]\rho+\frac{1}{2}[a^{\dagger}b-a b^{\dagger},\rho]-([a^{\dagger},b \rho]+[\rho b^{\dagger},a])
\end{equation*}
to convert this to explicit Lindblad form. For the two cases \(\Dc=\pm\om^{\mathrm{eff}}\) this leads to
\begin{subequations}
  \setlength\jot{1.5ex}
  \label{eq-ae:273}
  \begin{align}
    \dot{\bar{\rho}}&=
      \begin{multlined}[t]
        \cL_{\m} \bar{\rho}  + \ii \sqrt{\lambda_t\kappa_{\sc}\Gamma_+/4}[\ee^{-\ii \varphi_+}\cm\cc+\Hc{},\bar{\rho}]+\Gamma_-\lb[\cm]\bar{\rho}\\
        +(1-\lambda_t)\Gamma_+\lb[\cm^{\dagger}]\bar{\rho} +\lb[\sqrt{\kappa_{\sc}}\cc-\ii \ee^{\ii \varphi_+}\sqrt{\lambda_t\Gamma_+}\cm^{\dagger}]\bar{\rho},
      \end{multlined}\\
    \dot{\bar{\rho}}&=
      \begin{multlined}[t]
        \cL_{\m} \bar{\rho} + \ii \sqrt{\lambda_t\kappa_{\sc}\Gamma_-/4}[\ee^{-\ii \varphi_-}\cm^{\dagger}\cc+\Hc{},\bar{\rho}]+\Gamma_+\lb[\cm^{\dagger}]\bar{\rho}\\
        +(1-\lambda_t)\Gamma_-\lb[\cm]\bar{\rho} +\lb[\sqrt{\kappa_{\sc}}\cc-\ii \ee^{\ii \varphi_-}\sqrt{\lambda_t\Gamma_-}\cm]\bar{\rho},
      \end{multlined}
  \end{align}
\end{subequations}
respectively, with the definition $\varphi_{\pm}=\arg(\eta_{\pm})$. If we additionally choose \(\Dlc=\pm \om^{\mathrm{eff}}\) the resonant terms have phases \(\varphi_{\pm}=0\) and the resonant scattering rates are \(\Gamma_{\pm}=4g^2/\kappa_{\slc}\), while the off-resonant rates \(\Gamma_{\mp}=\epsilon \Gamma\) with \(\epsilon=1/[1+(4\om/\kappa_{\slc})^2]\). Equations \eqref{eq-ae:273} then lead to \eqref{eq:17} and \eqref{eq:18}.

\section{Evaluation of the correlation parameter $S$}
\label{sec:gaussian-dynamics}

To evaluate the quantity $S$ in \eqref{eq.BellIneq} for the bipartite system consisting of the two light pulses, we need to model the sequential measurement of the pulses using two microwave cavities containing a qubit.
The cavity modes then effectively constitute systems $A$ and $B$. Assuming that initially the mechanical mode is in a thermal state with a mean occupation number $n_0$ and both cavities are in the vacuum state, this allows us to find the final state $\rho_{\!AB}$ of modes $A$ and $B$ before the qubit measurements. This is achieved by first integrating \eqref{eq:17} for a duration $\tau_1$ and then \eqref{eq:18} for a duration $\tau_2$. The state $\rho_{\!AB}$ is then used to evaluate $S$, while the mechanical mode is traced out. (In fact the mechanics should nearly factor out from the rest of the system at this point.)

Under the dynamics described by the adiabatic master equations \eqref{eq:17} and \eqref{eq:18} the covariance matrix $\Sigma_{kl}=\frac{1}{2}\mean{X_kX_l+X_lX_k}-\mean{X_k}\mean{X_l}$ of the vector $\vc{X}=(x_a,y_a,x_b,y_b)$ evolves as described by the differential Lyapunov equation \(\dot{\mat{\Sigma}}=\mat{F}\mat{\Sigma}+\mat{\Sigma}\mat{F}^{\trans}+\mat{N}\). For the blue-detuned pulse we find%
  \onecolumngrid
    \begin{align*}
      \mat{F}&=
        \frac{1}{2}\begin{pmatrix}
          (1-\epsilon)\Gamma_{\mathrm{sq}}-\ga  & 0 & 0 & 0 \\
          0 & (1-\epsilon)\Gamma_{\mathrm{sq}}-\ga & 0 & 0 \\
          -2\sqrt{ \eta_{\mathrm{t}}
            \kappa_{\sc}\Gamma_{\mathrm{sq}}} & 0 & -\kappa_{\sc} & 0 \\
          0 & 2\sqrt{ \eta_{\mathrm{t}}
            \kappa_{\sc}\Gamma_{\mathrm{sq}}} & 0 & -\kappa_{\sc} \\
        \end{pmatrix},&
          \mat{N}&=\frac{1}{2}
            \begin{pmatrix}
              \tilde{\Gamma}_{\mathrm{sq}} & 0 & -\sqrt{ \eta_{\mathrm{t}}
                \kappa_{\sc}\Gamma_{\mathrm{sq}}} & 0 \\
              0 & \tilde{\Gamma}_{\mathrm{sq}} & 0 & \sqrt{ \eta_{\mathrm{t}}
                \kappa_{\sc}\Gamma_{\mathrm{sq}}} \\
              -\sqrt{ \eta_{\mathrm{t}}
                \kappa_{\sc}\Gamma_{\mathrm{sq}}} & 0 & \kappa_{\sc} & 0 \\
              0 & \sqrt{ \eta_{\mathrm{t}}
                \kappa_{\sc}\Gamma_{\mathrm{sq}}} & 0 & \kappa_{\sc} \\
            \end{pmatrix},
    \end{align*}
    while for the red-detuned second pulse we have
    \begin{align*}
      \mat{F}&=
        \frac{1}{2}\begin{pmatrix}
          -(1-\epsilon)\Gamma_{\mathrm{bs}}-\ga  & 0 & 0 & 0 \\
          0 & -(1-\epsilon)\Gamma_{\mathrm{bs}}-\ga & 0 & 0 \\
          -2\sqrt{ \eta_{\mathrm{t}}
            \kappa_{\sc}\Gamma_{\mathrm{bs}}} & 0 & -\kappa_{\sc} & 0 \\
          0 & -2\sqrt{ \eta_{\mathrm{t}}
            \kappa_{\sc}\Gamma_{\mathrm{bs}}} & 0 & -\kappa_{\sc} \\
        \end{pmatrix},&
          \mat{N}=\frac{1}{2}
          \begin{pmatrix}
            \tilde{\Gamma}_{\mathrm{bs}} & 0 & \sqrt{ \eta_{\mathrm{t}}
              \kappa_{\sc}\Gamma_{\mathrm{bs}}} & 0 \\
            0 & \tilde{\Gamma}_{\mathrm{bs}} & 0 & \sqrt{ \eta_{\mathrm{t}}
              \kappa_{\sc}\Gamma_{\mathrm{bs}}} \\
            \sqrt{ \eta_{\mathrm{t}}
              \kappa_{\sc}\Gamma_{\mathrm{bs}}} & 0 & \kappa_{\sc} & 0 \\
            0 & \sqrt{ \eta_{\mathrm{t}}
              \kappa_{\sc}\Gamma_{\mathrm{bs}}} & 0 & \kappa_{\sc} \\
          \end{pmatrix},
    \end{align*}%
\twocolumngrid%
\noindent
with $\tilde{\Gamma}_i=(1+\epsilon)\Gamma_i +\gamma_m(2\bar{n} + 1)$. To find $S$ it suffices to find evaluate the covariance matrix at the end of the pulse sequence.

\begin{acknowledgments}
  S.\,G.\ H. thanks Joshua A.\, Slater for useful comments on the manuscript and Marissa Giustina for helpful discussions. We acknowledge support by the European Commission (SIQS, iQOEMS, ITN cQOM), the Austrian Science Fund (FWF): project number [F40] (SFB FOQUS), the Vienna Science and Technology Fund (WWTF) under Project ICT12-049, the Centre for Quantum Engineering and Space-Time Research (QUEST), the Gordon and Betty Moore Foundation and National Science Foundation under Grant Number 1125844. S.\,G.\ H. is supported by the Austrian Science Fund (FWF): project number [W1210] (CoQuS).
\end{acknowledgments}
\bibliography{BellInequality,BellInequalityOptomechanics,SMRef,Klemens}

\end{document}